\documentclass[mathleft]{an}
\usepackage{graphicx}
\usepackage{times}
\begin{document}

\Pagespan{789}{}
\Yearpublication{2006}
\Yearsubmission{2005}
\Month{11}
\Volume{999}
\Issue{88}

\title{More frequencies of KUV 02464+3239}

\author{Zs. Bogn\'ar\inst{1}\fnmsep\thanks{\email{bognar@konkoly.hu}\newline}, M. Papar\'o\inst{1}, A. M\'ar\inst{2}, Gy. Kerekes\inst{2}, P. P\'apics\inst{2}, L. Moln\'ar\inst{1,2}, E. Plachy\inst{2}, N. Sztank\'o\inst{2}, E. Bokor\inst{2}}
\titlerunning{More frequencies of KUV 02464+3239}
\authorrunning{Zs. Bogn\'ar}
\institute{
Konkoly Observatory of the Hungarian Academy of Sciences, P.O. Box 67., H-1525 Budapest, Hungary
\and 
E\"otv\"os Lor\'and University Faculty of Science, P\'azm\'any P\'eter s\'et\'any 1/A, H-1117 Budapest, Hungary}

\received{30 May 2005}
\accepted{11 Nov 2005}
\publonline{later}


\abstract{Preliminary results on KUV 02464+3239, a pulsating DA white dwarf are presented. Located near the red edge of the DAV instability strip, KUV 02464+3239 shows large amplitude and long period pulsation modes. Up to now only one mode was known from a 50-minute-long light curve. Our more extended observations allowed the identification of three additional frequencies. The presence of previously known harmonics were confirmed and weak subharmonics are also noticeable at some parts of the light curve. This suggests the dominance of nonlinear pulsation effects from time to time.}

\maketitle

\section{Introduction}
KUV 02464+3239 (WD 0246+326, $V$ $=$ 15.8) was discovered to be a luminosity variable DA white dwarf by Fontaine et al.~(2001) in 1999. Based on their 50-minute-long observation (hereafter CFHT data) they had found that the star's Fourier spectrum was dominated by a large amplitude peak at $\sim$832s. The non-sinusoidal shape of the light curve, and its similarity to GD 154's, were obvious. Accordingly, the harmonics of the mean peak were also present. Up to now, only these results on KUV 02464+3239 have been published.

As DAV stars cool, their convective zones deepen and more and more modes are driven. Towards the red edge of the DAV instability strip, besides the general trend of increasing pulsation periods and amplitudes with decreasing temperature, we see more complex light curves. Amplitude variations, non-sinusoidal light curves and Fourier spectra with harmonics and linear combinations of frequencies are common. These features show that the nonlinear pulsation effects become more dominant (Dolez et al.~2006 and references therein). The presence of subharmonics could be a sign of the route to chaos (Goupil, Auvergne \& Baglin 1988; Vauclair et al.~1989). 

\section{Observations and data reduction}

Preliminary results of our first observations -- 5 nights (12 hours) over 20 days interval -- are presented here. The observations were obtained in white light at Piszk\'estet\H o, the mountain station of Konkoly Observatory, with a Princeton Instruments VersArray:1300B CCD camera attached to the 1m RCC telescope. The journal of observations is given in Table~\ref{JoO}.

The photometric reductions were carried out by using standard IRAF\footnote{IRAF is distributed by the National Optical Astronomy Observatories, which are operated by the Association of Universities for Research in Astronomy, Inc., under cooperative agreement with the National Science Foundation.} packages. Data analyis was made by use of the MuFrAn (Multi-Frequency Analyzer) package (Koll\'ath 1990; Csubry \& Koll\'ath 2004). MuFrAn provides efficient tools for frequency determination with its standard analyzer applications (FFT, DFT, linear and nonlinear fitting options) and graphics display routines. The program handles un\-e\-qual\-ly spaced and gapped observational data.

\begin{table}
\caption{Journal of Observations on KUV 02464+3239}
\label{JoO}
\begin{tabular}{cccccc}\hline
Date & Start Time & Length &  Total Points\\ 
(UT) & (HJD 2450000+)  & (hrs) &  \\
\hline
2006 Oct 06 & 4014.577 & 2.0 & 250 \\
2006 Oct 07 & 4015.584 & 1.9 & 560 \\
2006 Oct 09 & 4017.546 & 2.4 & 650 \\
2006 Oct 11 & 4019.543 & 3.0 & 800 \\
2006 Oct 25 & 4034.283 & 1.2 & 130 \\
2006 Oct 26 & 4034.460 & 2.0 & 494 \\
\hline
Total obs. time: & & 12.5 &\\
\hline
\end{tabular}
\end{table}

\section{Frequency analysis}

\begin{figure*}
\begin{centering}
\includegraphics[width=70mm]{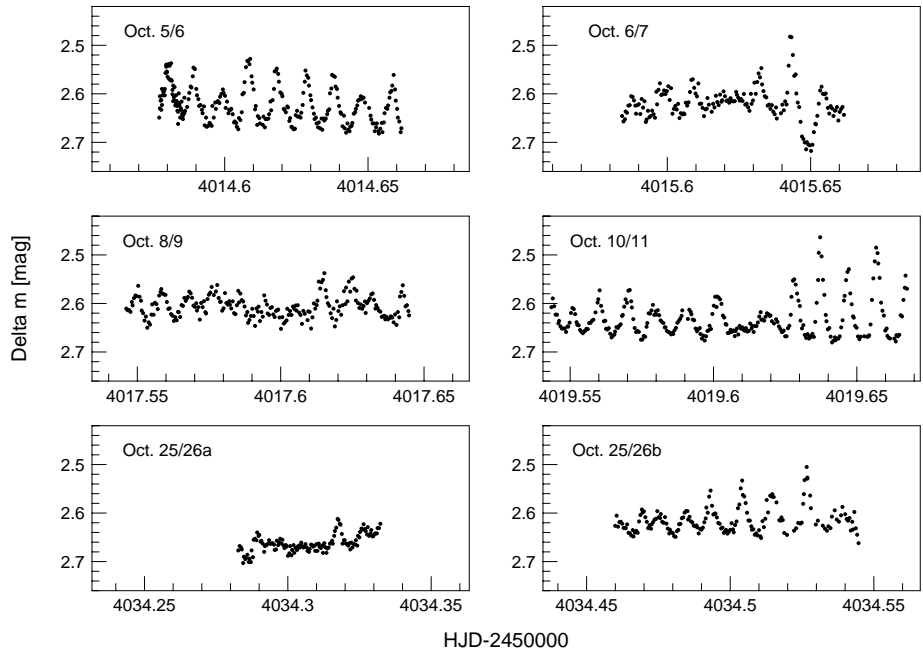}
\caption{Light curve of KUV 02464+3239 observed at Konkoly Observatory in 2006 October. The differential magnitudes with respect to the comparison star GSC2.2 N3333323300 are plotted as a function of Heliocentric JD. The points represent a sampling time of 30s.}
\label{lightcurves}
\end{centering}
\end{figure*}

In accordance with its position near the red edge of the DAV instability strip, KUV 02464+3239 ($T_{eff}$ $=$ 11290K, log $g$ $=$ 8.08, Fontaine et al.~2001) shows large amplitude and long period pulsation modes. The whole light curve is plotted in Fig.~\ref{lightcurves}, the points represent a sampling time of 30s. The individual cycles show rapid rises with sharp maxima and broader minima, which suggest presence of harmonics with significant power. The unusual deep minima at JD 4015.65 needs more investigation. Alternation of cycles with small and large maxima are also conspicuous at JD 4019.62-4019.67 and at JD 4034.49-4034.53.

\begin{figure}
\begin{centering}
\includegraphics[width=80mm]{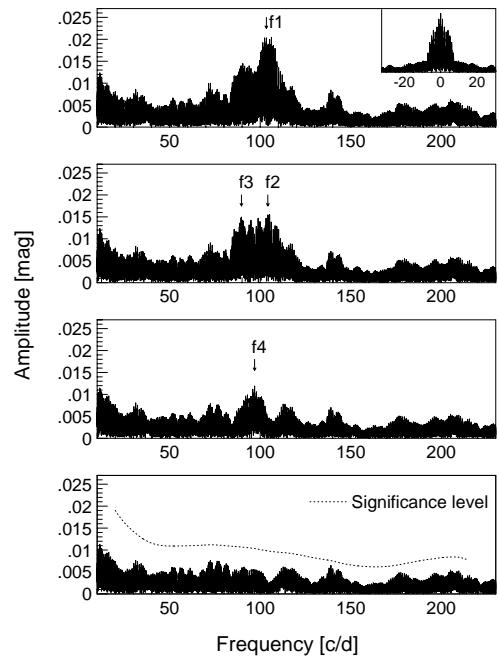}
\caption{Frequency analysis of the whole data set. The panels show the successive prewhitening steps (from top to bottom); the window function is given in the insert. The dashed line corresponds to the significance level at signal-to-noise ratio of $\sim$4.0.}
\label{spectra}
\end{centering}
\end{figure}

The Fourier spectrum derived from the whole light curve, the spectral window and the consecutive prewhitening steps are given in Fig.~\ref{spectra}. The spectral window has a rather complex structure because of data gaps. The final window function of our complete observations will differ from this preliminary one's. The strong aliasing renders the identification of close frequencies like {\it f$_1$} and {\it f$_2$} more difficult, however, the structure of the spectrum around 104 c/d clearly shows the presence of two frequencies. The dashed line in the last panel indicates the significance level at signal-to-noise ratio of $\sim$4.0. Peaks below this level are considered not to be intrinsic to the star.

Period and amplitude values are given in Table~\ref{freq-ampl} together with the CFHT data. Our analysis shows that large amplitude peaks can be found in a relatively small frequency range between 96 and 104 c/d. Considering the period values of {\it f$_1$}, {\it f$_2$} and the CFHT data, a triplet structure is also possible, however, the relative amplitudes of the components change from time to time.

\begin{table}
\caption{Frequencies derived in KUV 02464+3239}
\label{freq-ampl}
\begin{tabular}{cccc|cc}\hline
\multicolumn{4}{c}{Konkoly Observatory 2006} & \multicolumn{2}{c}{CFHT 1999}\\
\hline
 & {\it f} & P & {\it A} & P & {\it A}\\
& (c/d) & (s) & (mmag) & (s) & (mmag)\\
\hline
{\it f$_1$} & 103.40 & 835.6 & 20.8 & 831.6 & 39\\
{\it f$_2$} & 104.19 & 829.3 & 17.8 & &\\
{\it f$_3$} & 89.65 & 963.7 & 16.4 & &\\
{\it f$_4$} & 96.90 & 891.6 & 12.5 & &\\
\hline
\end{tabular}
\end{table}

\begin{figure*}
\begin{centering}
\includegraphics[width=70mm]{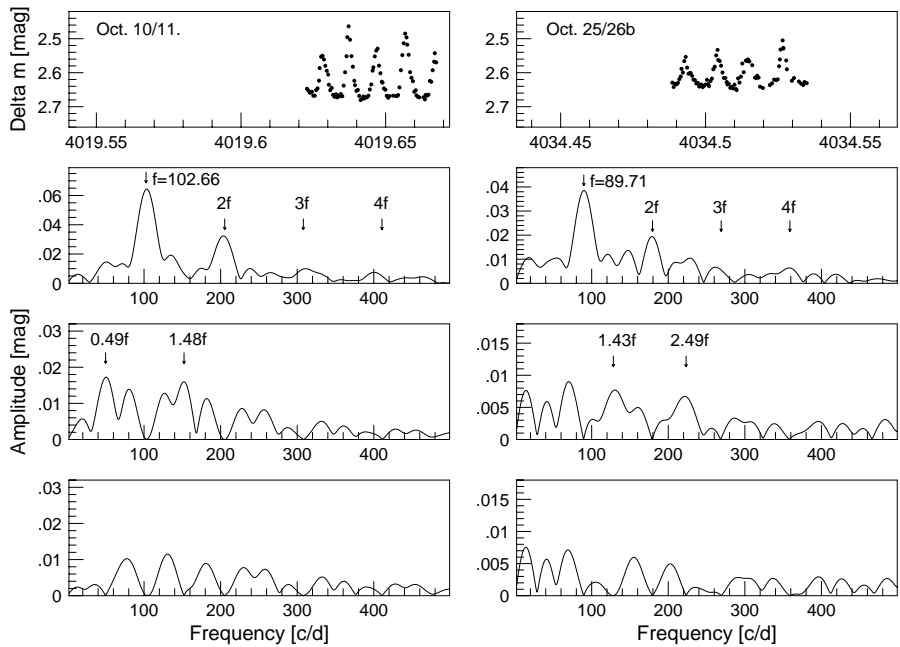}
\caption{Investigation of harmonics and subharmonics based on some parts of the light curve. Their Fourier spectra and the successive prewhitening steps are given in the panels (from top to bottom). Note the different vertical scales on the panels.}
\label{harmonics}
\end{centering}
\end{figure*}

We analysed separately the two parts of the light curve where alternating smaller and larger maxima were observed. The Fourier spectra derived from these segments of the light curve and the successive prewhitening steps are given in Fig.~\ref{harmonics}. These spectra are dominated by a mode near {\it f$_1$} or {\it f$_3$} and their second harmonics. Further harmonics and subharmonics are also noticeable with smaller amplitudes. This structure of the spectra reveals that the dynamics of the star is dominated by nonlinear processes. The subharmonics can be interpreted as a possible signature of the star's evolution towards a chaotic behaviour throught a cascade of period doubling bifurcations (Goupil et al.~1988; Vauclair et al.~1989).

GD 154, a photometric twin of KUV 02464+3239, has a Fourier spectrum with similar structure (Robinson et al. 1978). The DAV star G191-16 (Vauclair et al.~1989) and, from another spectral type, the pulsating DB white dwarf PG 1351+489 (Goupil et al.~1988) also show alternation of small and large peaks in their light curve from time to time. Accordingly, the presence of harmonics and subharmonics are also identified in their Fourier spectra.

\section{Conclusion}

Our preliminary results on the DAV type white dwarf KUV 02464+3239 disclosed its remarkable behaviour. The properties of the pulsation change on a few hours time scale; harmonics and subharmonics dominate the Fourier spectrum of the light curve at certain times, while there are intervals when these phenomena disappear. These features show the importance of nonlinear effects. The presence of subharmonics suggests that the star undergoes period doubling bifurcations. This could be a sign of the system's evolution toward chaos (Goupil et al.~1988; Vauclair et al.~1989).

Further observations are under analysis from the 2006-07 observing season. These data will help us to determine the pulsation frequencies more accurately, with special regard to nonlinear phenomena.


\end{document}